\documentclass[twocolumn]{aastex631}

\newcommand{\host}{HD\ {143761}}
\newcommand{\rcb}{$\rho\ \mathrm{CrB}$ }
\newcommand{\nobs}{163}
\newcommand{\ngoodobs}{153}
\newcommand{\nnights}{89}
\newcommand{\ngoodnights}{82}

\newcommand{\kepler}{{\em Kepler} \ }
\newcommand{\hundredearths}{{\em {100} \ Earths \ Survey}}
\newcommand{\expres}{EXPRES}

\newcommand{\ms}{\mbox{m s$^{-1}$}}

\newcommand{\mearth}{M$_{\earth}$}

\newcommand{\vsini}{$v \sin i$}

\newcommand{\teff}{${T_{\rm eff}}$}
\newcommand{\logg}{$\log{g}$}

\newcommand{\snr}{SNR}
 
\newcommand{\update}{}

\shorttitle{Temperate Neptune, Hot Super Earth}
\shortauthors{Brewer et al.}
\graphicspath{{./}{figures/}}

\begin{document}

\title{EXPRES IV: Two Additional Planets Orbiting $\rho$ Coronae Borealis Reveal Uncommon System Architecture}

\correspondingauthor{John M. Brewer} 
\email{jmbrewer@sfsu.edu}
\author[0000-0002-9873-1471]{John M. Brewer}

\affiliation{Department of Physics and Astronomy, San Francisco State University, 1600 Holloway Ave, San Francisco, CA 94132, USA}

\author[0000-0002-3852-3590]{Lily L. Zhao}
\affiliation{Center for Computational Astrophysics, Flatiron Institute, Simons Foundation, 162 Fifth Avenue, New York, NY 10010, USA}

\author[0000-0003-2221-0861]{Debra A. Fischer}
\affiliation{Department of Astronomy, Yale University, 52 Hillhouse Ave, New Haven, CT 06511, USA}

\author[0000-0002-9288-3482]{Rachael M. Roettenbacher} 
\affiliation{Department of Astronomy, Yale University, 52 Hillhouse Ave, New Haven, CT 06511, USA}  
\affiliation{Department of Astronomy, University of Michigan, 1085 S.\ University Ave., Ann Arbor, MI 48109, USA}

\author[0000-0003-4155-8513]{Gregory W. Henry} 
\affiliation{Center of Excellence in Information Systems, Tennessee State University, Nashville, TN 37209, USA}

\author[0000-0003-4450-0368]{Joe Llama}
\affiliation{Lowell Observatory, 1400 Mars Hill Rd, Flagstaff, AZ 86001, USA}

\author[0000-0002-4974-687X]{Andrew E. Szymkowiak}
\affiliation{Department of Astronomy, Yale University, 52 Hillhouse Ave, New Haven, CT 06511, USA}

\author[0000-0001-9749-6150]{Samuel H. C. Cabot}
\affiliation{Department of Astronomy, Yale University, 52 Hillhouse Ave, New Haven, CT 06511, USA}

\author[0000-0002-5870-8488]{Sam A. Weiss}
\affiliation{Department of Astronomy, Yale University, 52 Hillhouse Ave, New Haven, CT 06511, USA}

\author[0000-0001-5387-9053]{Chris McCarthy}
\affiliation{Department of Physics and Astronomy, San Francisco State University, 1600 Holloway Ave, San Francisco, CA 94132, USA}



\begin{abstract}

 Thousands of exoplanet detections have been made over the last twenty-five years using Doppler observations, transit photometry, direct imaging, and astrometry. Each of these methods is sensitive to different ranges of orbital separations and planetary radii (or masses). This makes it difficult to fully characterize exoplanet architectures and to place our solar system in context with the wealth of discoveries that have been made. Here, we use the EXtreme PREcision Spectrograph (EXPRES) to reveal planets in previously undetectable regions of the mass-period parameter space for the star $\rho$ Coronae Borealis. We add two new planets to the previously known system with one hot Jupiter in a 39-day orbit and a warm super-Neptune in a 102-day orbit. The new detections include a temperate Neptune planet ($M{\sin{i}} \sim 20$ M$_\oplus$) in a 281.4-day orbit  and a hot super-Earth ($M{\sin{i}} = 3.7$ M$_\oplus$) in a 12.95-day orbit. This result shows that details of planetary system architectures have been hiding just below our previous detection limits; this signals an exciting era for the next generation of extreme precision spectrographs. 

\end{abstract}




\section{Introduction} \label{sec:intro}
\update{The detection of 51 Peg b \citep{1995Natur.378..355M} was enabled by the ground-breaking radial velocity (RV) precision of $\sim15$ \ms\ \citep{10.1086/166608,10.1006/icar.1995.1130}. The RV technique continued to improve and soon reached a new state of the art precision of $\sim$3 \ms\ \citep{10.1086/133755}. Several additional gas giant planets in short-period orbits were detected and the seventh of these was a Jupiter-mass planet in a 39.6-day orbit around \rcb\ b \citep{10.1086/310754}.}

\update{In the ensuing years, lower mass planets in wider orbits were detected. From the earliest days of exoplanet surveys, it was understood that obscuring RV scatter was correlated with strong chromospheric activity \citep{1991LNP...390...19C, Queloz2001}. However, it it was not clear whether the ultimate precision for radial velocity measurements would be set by instrumental stability or velocity noise from the stellar photosphere.}

\update{The High Angular Resolution Planetary Spectrograph (HARPS) achieved the next break-through in RV precision when it reached a single measurement precision of 1 \ms\ \citep{Queloz2001,Mayor2003,Pepe2003}. 
This precision improvement and a relatively high-cadence program made a big difference in the ability to detect lower amplitude and multi-planet signals, epitomized by the addition of a sub-Neptune planet with a velocity amplitude of only 4 \ms\ to the $\mu$ Arae system \citep{10.1086/321467, 2004ApJ...617..575M, 10.1051/0004-6361:200400076, 2007A&A...462..769P}.}

\update{The latest generation of spectrographs, including ESPRESSO \citep{10.1051/0004-6361/202038306}, EXPRES \citep{2016SPIE.9908E..6TJ}, and NEID \citep{10.1117/12.2234411} have established a new state-of-the-art RV precision of 0.3 \ms. The story of \rcb\ illustrates the significance of this increased precision and high observing cadence.}

\citet{10.1086/310754} identfied a 39.6 day, 67 \ms sinusoidal signal around \rcb in only 41 nights of observations over one observing season. This planet was well within the capabilities of all of the planet-hunting spectrographs of the time.  The Hamilton spectrograph at Lick Observatory had been upgraded to 3 \ms\ precision in 1994, and \rcb\ was added to the program in 1997.  Over the next 10 years an additional 26 of these higher precision measurements had been obtained, refining the orbital parameters slightly but identifying no new signals \citep{Butler:2006dd}. The RMS scatter to the one-planet fit was 6.9 \ms, and the `stellar jitter' was estimated to be 3.9 \ms.  

By the time of the \citet{Butler:2006dd} follow-up on \rcb there were multiple spectrographs with $\sim 1-3$ \ms\ measurement precision \citep[][ and references therein]{Fischer:2016hl}. \rcb was added to the target list at Keck/HIRES and then the Automated Planet Finder (APF) as part of the Eta-Earth RV Survey \citep{2009ApJ...696...75H}. \citet{10.3847/0004-637x/830/1/46} used 519 Keck/HIRES measurements and 157 APF velocities, taken over 8 years, to identify a second planet.  Despite their long time baseline, the Whipple and Lick velocities were omitted; their comparatively low cadence and low precision did not contribute to the significance of the detection.  This second planet, \rcb\ c, was found to be a 25 \mearth\ planet on a 102-day orbit with an RV semi-amplitude of 3.74 \ms \citep{10.3847/0004-637x/830/1/46}, a signal smaller than the `stellar jitter' reported a decade earlier.  The revised estimate for stellar jitter by \citet{10.3847/0004-637x/830/1/46} was 2.57 \ms.  They saw no evidence of additional planets, but mentioned possible structure to the residuals with a period $\gtrsim 10$ years.

With every improvement in instrumental precision, our estimate of the `stellar noise floor' has changed.  The residual velocity scatter attributed to stellar jitter has proven to be a combination of unknown instrumental errors, stellar variability, and multiple planetary signals \citep{Isaacson:2010gk,10.3847/1538-3881/ab855a}.  With the new generation of extreme precision radial velocity (EPRV) spectrographs \citep{2016SPIE.9908E..6TJ,10.1051/0004-6361/202038306,10.1117/12.2234411}, the instrumental contribution is often smaller than the stellar component \citep{Isaacson:2010gk,10.3847/1538-3881/ab855a}. \update{We now know that nearly all stars have planets \citep{Howard:2010es,Lovis2011,Mayor:2011tb,2015ApJ...809....8B,Hsu:2019jf}.} However, our knowledge about the detailed planetary architectures are limited by our lack of sensitivity to a large portion of the mass-period (or radius-period) parameter space. Current statistics suggest that there are an average of three planets per stellar host and that the average planet size is between that of Earth and Neptune \citep[][and references therein]{10.3847/0004-637x/828/2/99,10.3847/1538-3881/abab0b,Hsu:2019jf,ForemanMackey:2014bj,2016AJ....152..206F,2015ARA&A..53..409W,10.1007/978-3-319-30648-3_195-1,2011ApJ...736...19B,Hsu:2019jf,2021arXiv210302127Z}.

Planet multiplicity greatly increases the number of observations needed to disentangle the Keplerian signals \citep{2021arXiv210504703H}.  If those are obtained over long timescales, then different phases in the activity cycle of the star are sampled, making it more difficult to detect low amplitude signals.  Combining high cadence with high instrumental precision can help us identify the small signals that may be lurking in our data.

Radial velocity data has been collected for hundreds of stars for more than 25 years \citep[e.g.][]{10.1086/321467,1993ASPC...36..267C,10.1007/10720961_79,10.1051/0004-6361:20040389,2010ApJ...721.1467H,2014ApJS..210....5F}.  For the last 10 years, the precision of these surveys has been $\sim 1$ m/s \citep{Fischer:2016hl}.  This precision, combined with relatively low observing cadence \update{in most surveys} has made it challenging to detect planets with $M < 30 M_\oplus$ on periods beyond $\sim 100$ days.  For systems with planets on orbits shorter than 40 days, we have still been limited to planets of a few Earth masses for G and K dwarfs, missing many of the smaller planets that have been identified by transit surveys.  It is more difficult to disentangle multi-planet signals, especially when one or more are close to the instrumental precision \citep{2021arXiv210504703H}, preventing detection of architectures similar to the solar system.

The \hundredearths\ is a high-cadence EPRV survey \citep{10.3847/1538-3881/ab99c9} designed to locate the small and intermediate mass planets in orbits that have so far eluded both transit and RV surveys. It specifically aims to detect planets in the habitable zones of Sun-like (G and K dwarf) stars.  The program combines the extreme stability of the EXPRES spectrograph \citep{2016SPIE.9908E..6TJ,2020AJ....159..238B,2020AJ....159..187P} with very high cadence to identify low amplitude signals in multi-planet systems on relatively short timescales.  The rapid switching capabilities of the Lowell Discovery Telescope \citep[LDT,][]{Levine:2012cr,DeGroff:2014hf} allows observing in quarter night increments, and has even been used for very high cadence observations with allocations of just 30 minutes.  \citet{10.3847/1538-3881/ab99c9} showed that for HD 3651 it was possible to recover the known planet parameters with residuals of 58 cm/s using only 60 observations taken over $\sim 5$ months, roughly two orbital periods of HD 3651b. Here, we will present the first new detections by this program: two additional planets around \rcb (\host). The architecture of the four planet system differs from most previously detected systems in arrangment and variety of masses.

\section{Observations} \label{sec:observations}

Since August 2019 the \hundredearths\ has collected \nobs\ observations of \host\ on \nnights\ separate nights (Table \ref{tab:143761vels}). We use an exposure meter to stop all observations at the same \snr\ in order to reduce the effects of charge transfer inefficiency (CTI) on the radial velocities\update{; at \snr=300, CTI contributes about 10 c\ms offset in the measured RV \citep{2020AJ....159..238B}}. When the program began, we obtained 3 consecutive observations for \rcb, each with \snr=250 at 500 nm.  This strategy allowed us to evaluate the short-term RV scatter in our data for our target stars.  This cadence was updated in 2020 with 2 consecutive observations at \snr=310, improving the duty cycle of our observations while maintaining the same nightly RV precision. Clouds and bad seeing sometimes limit the number of exposures we can obtain, and we stop all exposures at 1200s to ensure a reasonable correction for the barycentric velocity\update{; integrations not reaching the required \snr\ in this time are discarded.} Five observations of \host\ during 2020 were obtained using the lower per observation \snr, while the remainder were at the higher value as well as higher cadence. We increased the observing priority for this star in 2021. All observations that met the lower \snr\ requirement in less than 20 minutes were included in this analysis\update{: \ngoodobs\ observations on \ngoodnights\ nights}. The spectra were extracted using a flat relative optimal extraction, and radial velocities were derived using forward modeling \citep{2020AJ....159..187P}.

\begin{deluxetable}{ccc}
\tablecaption{EXPRES RVs of HD 143761 \label{tab:143761vels}}
\tablehead{\colhead{BMJD} & \colhead{Vel \ms} & \colhead{Err \ms}}
\startdata
58983.23654927 & 28.913 & 0.420 \\
58983.23864388 & 29.743 & 0.438 \\
58983.24077284 & 27.107 & 0.445 \\
59012.37000357 & 52.129 & 0.445 \\
59012.37211914 & 54.251 & 0.445 \\
59012.37400430 & 54.915 & 0.458 \\
59017.30889883 & 59.608 & 0.419 \\
59335.41151111 & 58.781 & 0.377 \\
59335.41464020 & 58.331 & 0.378 \\
59335.41764142 & 58.228 & 0.372
\enddata
\end{deluxetable}

During the night, we obtain laser frequency comb (LFC) observations every 20-30 minutes and Thorium-Argon emission lamp (ThAr) observations every hour.  Our pipeline \citep{2020AJ....159..187P} uses the ThAr observations for a broad wavelength solution over the entire spectrum and uses the cross-correlation technique to derive an absolute radial velocity.  We also derive a separate, more precise, wavelength solution using the LFC frames between $\sim 5000$~\AA\ and 7200~\AA\ to construct a hierarchical, non-parametric model \citep{10.3847/1538-3881/abd105}. With these wavelengths, we use forward modelling of individual $\sim 2$~\AA\ chunks to derive precise relative velocities with single-measurement precision of 30 cm/s at \snr$= 250$ \citep{10.3847/1538-3881/ab99c9,2020AJ....159..187P}.

\subsection{Stellar Properties} \label{sec:stellar_properties}

\begin{deluxetable}{cc}
\tablecaption{Stellar Properties for HD 143761 \label{tab:stellar_props}}
\tablehead{\colhead{Parameter} & \colhead{Value}}
\startdata
Identifier & HD 143761 \\
T$_{\mathrm{eff}}$ [K] & 5817 $\pm 24$ \\
$\log{g}$ [m s$^{-2}$] & 4.25 $\pm 0.05$ \\
$v \sin{i}$ [km s$^{-1}$] & 0.8 $\pm 0.3$ \\
$V$ [mag] & 5.39\tablenotemark{a} \\
$B - V$ & 0.61\tablenotemark{a} \\
$BP$ [mag] & 5.55 $\pm 0.05$\tablenotemark{b}\\
$RP$ [mag] & 4.75 $\pm 0.05$\tablenotemark{b} \\
Distance [pc] & 17.497 $\pm 0.015$\tablenotemark{b} \\
{[C/H]}	&	-0.14 $\pm 0.03$ \\
{[N/H]}	&	-0.29 $\pm 0.04$ \\
{[O/H]}	&	+0.07 $\pm 0.04$ \\
{[Na/H]}	&	-0.24 $\pm 0.02$ \\
{[Mg/H]}	&	-0.11 $\pm 0.01$ \\
{[Al/H]}	&	-0.07 $\pm 0.03$ \\
{[Si/H]}	&	-0.16 $\pm 0.01$ \\
{[Ca/H]}	&	-0.12 $\pm 0.02$ \\
{[Ti/H]}	&	-0.08 $\pm 0.01$ \\
{[V/H]}	&	-0.15 $\pm 0.03$ \\
{[Cr/H]}	&	-0.24 $\pm 0.02$ \\
{[Mn/H]}	&	-0.40 $\pm 0.02$ \\
{[Fe/H]}	&	-0.20 $\pm 0.01$ \\
{[Ni/H]}	&	-0.22 $\pm 0.01$ \\
{[Y/H]}	&	-0.25 $\pm 0.03$ \\
Mass [M$_\odot$] & 0.95 $\pm 0.01$ \\
Luminosity [L$_\odot$] & 1.82 $\pm 0.08$ \\
Radius [R$_\odot$] & 1.34 $\pm 0.04$ \\
Age [Gyrs] & 10.2 $\pm 0.5$ \\
\enddata
\tablerefs{\tablenotetext{a}{\citet{vanLeeuwen:2007dc}} \tablenotetext{b}{\citet{10.1051/0004-6361/202039709}}}
\end{deluxetable}

Following the procedure of \citet{Brewer:2016gf}, we use Spectroscopy Made Easy (SME) to forward model $\sim 350$~\AA\ of the spectrum to determine effective temperature, surface gravity, projected rotational velocity, and overall metallicity along with precise abundances for 15 elements (C, N, O, Na, Mg, Al, Si, Ca, Ti, V, Cr, Mn, Fe, Ni, and Y; Table \ref{tab:stellar_props}).  Some RV surveys use an iodine cell for wavelength calibration on each science frame, prohibiting use of those observations for abundance analysis. Because the wavelength calibration for EXPRES is performed in frames adjacent to our stellar observations, we are able to analyze every observation to constrain the uncertainties in the derived parameters.

After deriving the stellar parameters, we combine the \teff, \logg, [Fe/H], and [$\alpha$/Fe] measurements with the $V$ magnitude, $B - V$ color, and Gaia DR3 distance measurement \citep{10.1051/0004-6361/202039834} to derive the stellar luminosity and radius (Table \ref{tab:stellar_props}).  We use those same inputs as prior constraints in our isochrone analysis to fit for the mass and age.  In addition to the Yonsei-Yale isochrone \citep[Y2 isochrones;][]{2004ApJS..155..667D} analysis as described in \citet{Brewer:2016gf}, we performed a more detailed analysis for the sake of comparison with the stellar properties from the previous planet detection around \rcb\ \citep{10.3847/0004-637x/830/1/46}.

Although most of the stellar properties are consistent with those reported in earlier studies, there are a couple of items of interest in the context of its planets.  As an early system with a hot Jupiter, \rcb\ stands out with its relatively low metallicity of [Fe/H] = -0.2.  This makes it a rarity in the context of the giant planet metallicity correlation \citep{2005ApJ...622.1102F}.  The abundance pattern itself is also interesting, with oxygen significantly elevated compared to scaled Solar values.  However, an investigation into possible causes for this lies outside the scope of this paper.

In addition to having sub-Solar metallicity, the star is evolved. It is an early sub-giant with a radius of 1.34 $R_\odot$ and near Solar mass. However, there is slight discrepency in the \teff, [Fe/H], and mass between our analysis and those used in \citet{10.3847/0004-637x/830/1/46} based on the interferometric measurements of \citet{10.1093/mnras/stt2360}.  In that analysis, they find the \teff\ is $\sim 200$ K cooler, it is 0.1 dex more metal poor, and the mass of $0.89 \ M_\odot$  $6 - 10\%$ smaller.  That temperature falls outside the distribution of literature values, which have a mean of $5798 \pm 58$ K that is consistent with our value.

\subsection{Isochrone Analysis}

\begin{figure}[htb]
\centering
    \includegraphics[width=\linewidth]{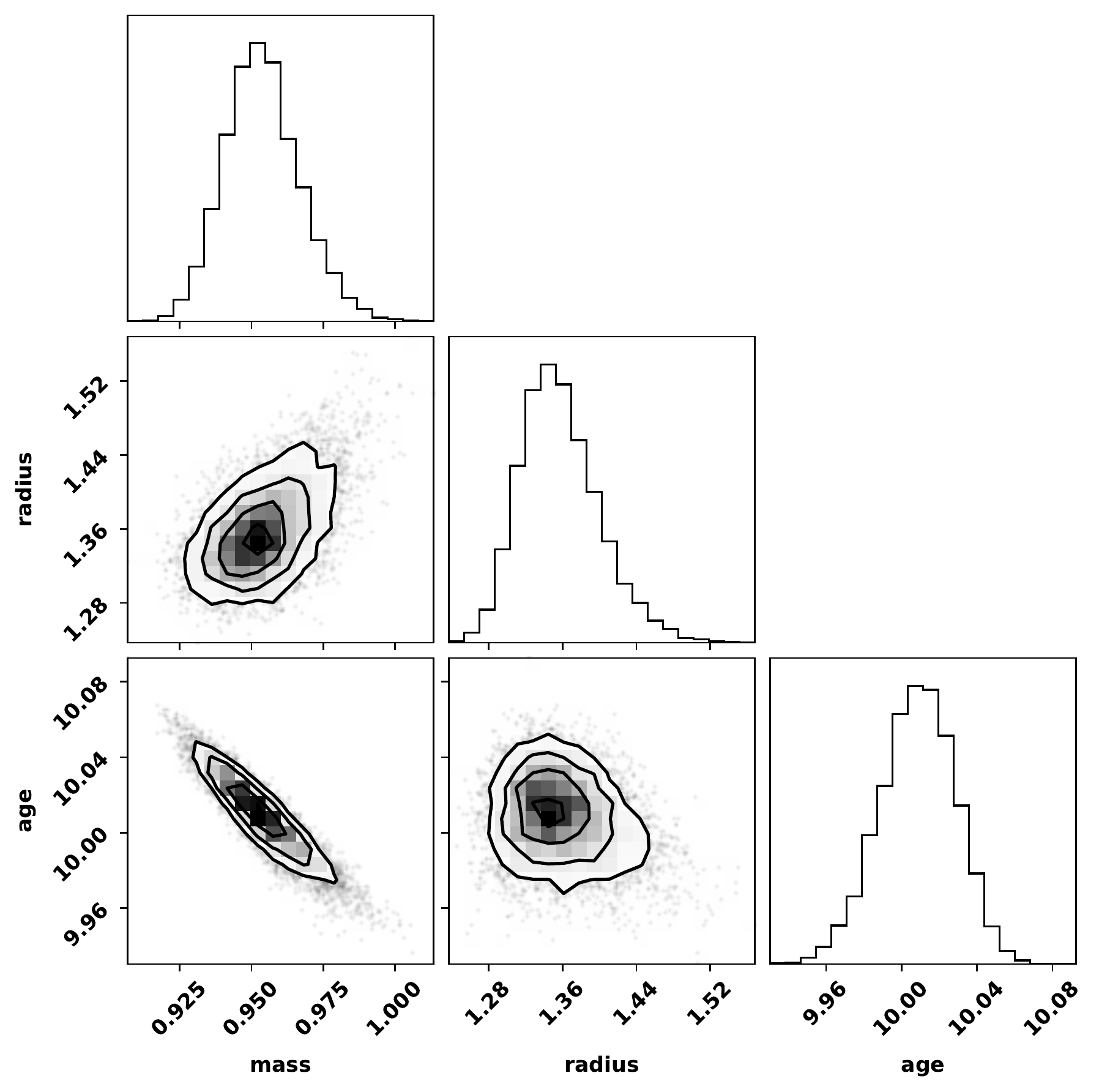}
\caption{Posterior distributions of stellar parameters (mass, radius, and $\log_{10}{\textrm{age}}$) fit using \texttt{isochrones} Python package show smooth distributions and are in good agreement with all of our observables.  The star is a 10.2 Gyr subgiant, but was a slightly sub-Solar metallicity 0.95 $M_\odot$ main sequence star.
\label{fig:isochrone_corner_plot}}
\end{figure}

Over its main sequence lifetime, elemental diffusion causes an apparent depletion in surface abundances \citep{2017ApJ...840...99D}. Determining the early main sequence properties of a turn-off star, where this depletion is at its maximum, requires a more careful analysis.  The \texttt{isochrones}\footnote{\url{http://github.com/timothydmorton/isochrones}} \citep{10.1088/0004-637x/809/1/25} Python package using the MIST isochrones \citep{Choi:2016wb} allows us to determine the initial parameters of the star and provides consistent evolutionary tracks to look at the main sequence behavior. Using our spectroscopic parameters combined with Gaia BP and RP colors and parallax, we found that the best fit was a $0.95 \pm 0.01$ $M_\odot$ at $10.2 \pm 0.5$ Gyr with initial [Fe/H] = -0.11 (Figure \ref{fig:isochrone_corner_plot}).  The mass and age are close to that derived from the Y2 isochrones ($0.98 \pm 0.01 M_\odot$ and $9.5 \pm 0.4$ Gyrs).  Both isochrone analyses derived a slightly lower \logg\ (4.16) than that derived via spectroscopy.

We then performed the same analysis using the parameters from \citet{10.3847/0004-637x/830/1/46} and find a mass of $M = 0.90 \pm 0.02 M_\odot$ at an age of $12.4^{+0.73}_{-0.52}$ Gyrs and an initial metallicity of [Fe/H] = -0.26.  \update{At this age, that would imply this star was in one of the first generations of stars in the Milky Way, and yet it is relatively metal rich.} Although this is possible, we find that our parameters cover a more reasonable age range and so adopt the parameters in Table \ref{tab:stellar_props} for the rest of this work.  The adopted distance from the prior work (closer by 0.25 pc) may have influenced their adoption of parameters for a smaller, cooler star. \update{The $\sim 5\%$ higher stellar mass will result in a 4\% increase in the planet minimum masses.}

\section{Doppler Analysis} \label{sec:analysis}

We use a Lomb-Scargle periodogram to identify the strongest period in the velocities with a false alarm probability (FAP) below $10^{-3}$. We fit and subtract a Keplerian signal with that period then repeat the procedure until there are no periods with FAP $< 10^{-3}$. Using the resulting parameters as starting values, we perform a simultaneous fit for the four Keplerian signals.  For the two new planets, we placed a Gaussian prior on the eccentricity centered at 0.05 with $\sigma = 0.05$ in keeping with the other two planets. \update{Adding a linear trend with a flat prior gave an improved fit with a trend of about -0.6 $ms^{-1}year^{-1}$. To test whether or not a stellar jitter term was warrented, we added jitter as a free parameter to our model with an initial value of 0.35 \ms, corresponding to the intra-night scatter for this star. The resulting model has nearly identical parameters to the previous best fit model with the exception of planet d, which has a period two days shorter. The reduced $\chi^2$ of this model was 1.3.} Model comparison analysis showed that a model with eccentricity fixed to zero for the longest period planet (d) was preferred over allowing it to vary.  This is likely due to the incomplete phase coverage in our data for this planet; the period is longer than an observing season.  

We use this model to set the priors on a Markov Chain Monte Carlo (MCMC) analysis using the python software package RadVel \citep{10.1088/1538-3873/aaaaa8} to obtain uncertainties on the parameters.  To keep the scales of the parameters comparable, we fit with some parameters combined and scaled: $\sqrt{e} \cos{\omega}$, $\sqrt{e} \sin{\omega}$ and $\log{K}$.  The MCMC analysis used 50 walkers and the first 300,000 steps of burn-in were discarded; the chains were well mixed after 1,920,000 steps. The posterior distributions were relatively tight and generally Gaussian in shape (Appendix \ref{appendix:Distributions}).

\begin{deluxetable*}{lrrr}
\label{tab:orbit_params}
\tablecaption{ MCMC Posteriors }
\tablehead{
  \colhead{Parameter} & 
  \colhead{Credible Interval} & 
  \colhead{Maximum Likelihood} & 
  \colhead{Units}
}
\startdata
\sidehead{\bf{Orbital Parameters}}
  $P_{b}$ & $39.8438\pm 0.0027$ & $39.844$ & days \\
  $T\rm{conj}_{b}$ & $55479.62\pm 0.29$ & $55479.6$ & JD \\
  $T\rm{peri}_{b}$ & $55498.7^{+0.75}_{-39.0}$ & $55499$ & JD \\
  $e_{b}$ & $0.038\pm 0.0025$ & $0.0379$ &  \\
  $\omega_{b}$ & $-1.577^{+0.06}_{-0.058}$ & $-1.577$ & radians \\
  $K_{b}$ & $67.28\pm 0.19$ & $67.28$ & m s$^{-1}$ \\
  $P_{c}$ & $102.19^{+0.27}_{-0.22}$ & $102.18$ & days \\
  $T\rm{conj}_{c}$ & $55629.3^{+8.6}_{-11.0}$ & $55629.5$ & JD \\
  $T\rm{peri}_{c}$ & $55609^{+13}_{-14}$ & $55609$ & JD \\
  $e_{c}$ & $0.096^{+0.053}_{-0.054}$ & $0.09$ &  \\
  $\omega_{c}$ & $0.17^{+0.44}_{-0.54}$ & $0.16$ & radians \\
  $K_{c}$ & $4.0^{+0.18}_{-0.17}$ & $4.02$ & m s$^{-1}$ \\
  $P_{d}$ & $282.2^{+2.2}_{-3.7}$ & $281.4$ & days \\
  $T\rm{conj}_{d}$ & $55583^{+54}_{-31}$ & $55596$ & JD \\
  $T\rm{peri}_{d}$ & $55512^{+55}_{-32}$ & $55525$ & JD \\
  $e_{d}$ & $\equiv0.0$ & $\equiv0.0$ &  \\
  $\omega_{d}$ & $\equiv0.0$ & $\equiv0.0$ & radians \\
  $K_{d}$ & $2.2\pm 0.14$ & $2.21$ & m s$^{-1}$ \\
  $P_{e}$ & $12.949\pm 0.014$ & $12.949$ & days \\
  $T\rm{conj}_{e}$ & $55498.5^{+4.6}_{-4.7}$ & $55498.3$ & JD \\
  $T\rm{peri}_{e}$ & $55496.2^{+5.7}_{-5.0}$ & $55495.4$ & JD \\
  $e_{e}$ & $0.126^{+0.054}_{-0.078}$ & $0.073$ &  \\
  $\omega_{e}$ & $-0.01^{+0.86}_{-1.0}$ & $-0.01$ & radians \\
  $K_{e}$ & $1.14\pm 0.015$ & $1.142$ & m s$^{-1}$ \\
\hline
\sidehead{\bf{Other Parameters}}
  $\gamma$ & $-0.74^{+0.32}_{-0.33}$ & $-0.75$ &  \ms \\
  $\dot{\gamma}$ & $-0.00146^{+0.00048}_{-0.00047}$ & $-0.00144$ & m s$^{-1}$ d$^{-1}$ \\
  $\ddot{\gamma}$ & $\equiv0.0$ & $\equiv0.0$ & m s$^{-1}$ d$^{-2}$ \\
  jitter & $1.111^{+0.067}_{-0.063}$ & $1.054$ &  \\
\enddata
\tablenotetext{}{1,620,000 links saved}
\tablenotetext{}{
  Reference epoch for $\gamma$,$\dot{\gamma}$,$\ddot{\gamma}$: 58982.73654927462
  Dates are Barycentric JD - 2,400,000
}
\label{tab:params}
\end{deluxetable*}

\begin{figure}
    \centering
    \includegraphics[width=\linewidth]{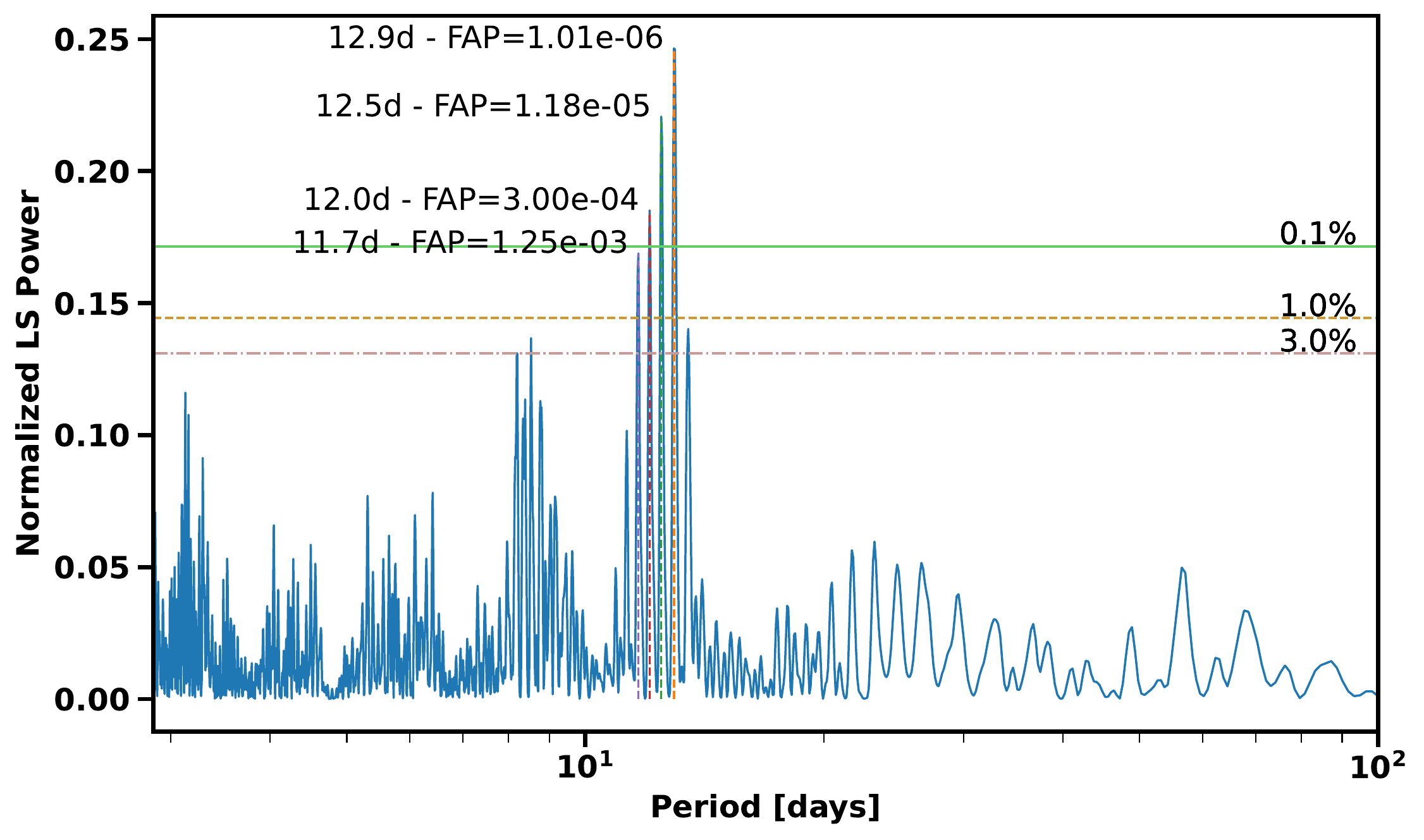}
    \caption{Periodogram to the residuals to a three-planet fit, with planets at 39.8, 102.2, and 283.8 days.  The highest peak (12.9 days) has an FAP=$1.0 \times 10^{-6}$, indicating a possible fourth planet.  The four highest peaks are labelled in the plot along with the 3\%, 1\%, and 0.1\% FAP levels.  No significant power was found at longer periods, which are not shown.  A periodogram of the residuals to a four planet fit showed no significant peaks.}
    \label{fig:3p_periodogram}
\end{figure}

\update{The highest peak in the residuals to a three-planet fit is at 12.9 days and has an FAP=$1.0 \times 10^{-6}$ (Figure \ref{fig:3p_periodogram}).  After adding a fourth Keplerian at 12.9 days, there are no significant peaks in the periodogram of the residuals.}

\update{We used $\Delta BIC$ and $\Delta AIC$ to compare two, three, and four-planet models.  In each case, the model with the additional planet was overwhelmingly preferred over the simpler system. Our final four-planet fit (Table \ref{tab:orbit_params}) has a $\Delta BIC_{3p-4p} = 13$ compared to the best three-planet fit.} The derived planetary masses and semi-major axes can be found in Table \ref{tab:derived_params}.

\begin{figure}[ht]
\centering
    \includegraphics[width=\linewidth]{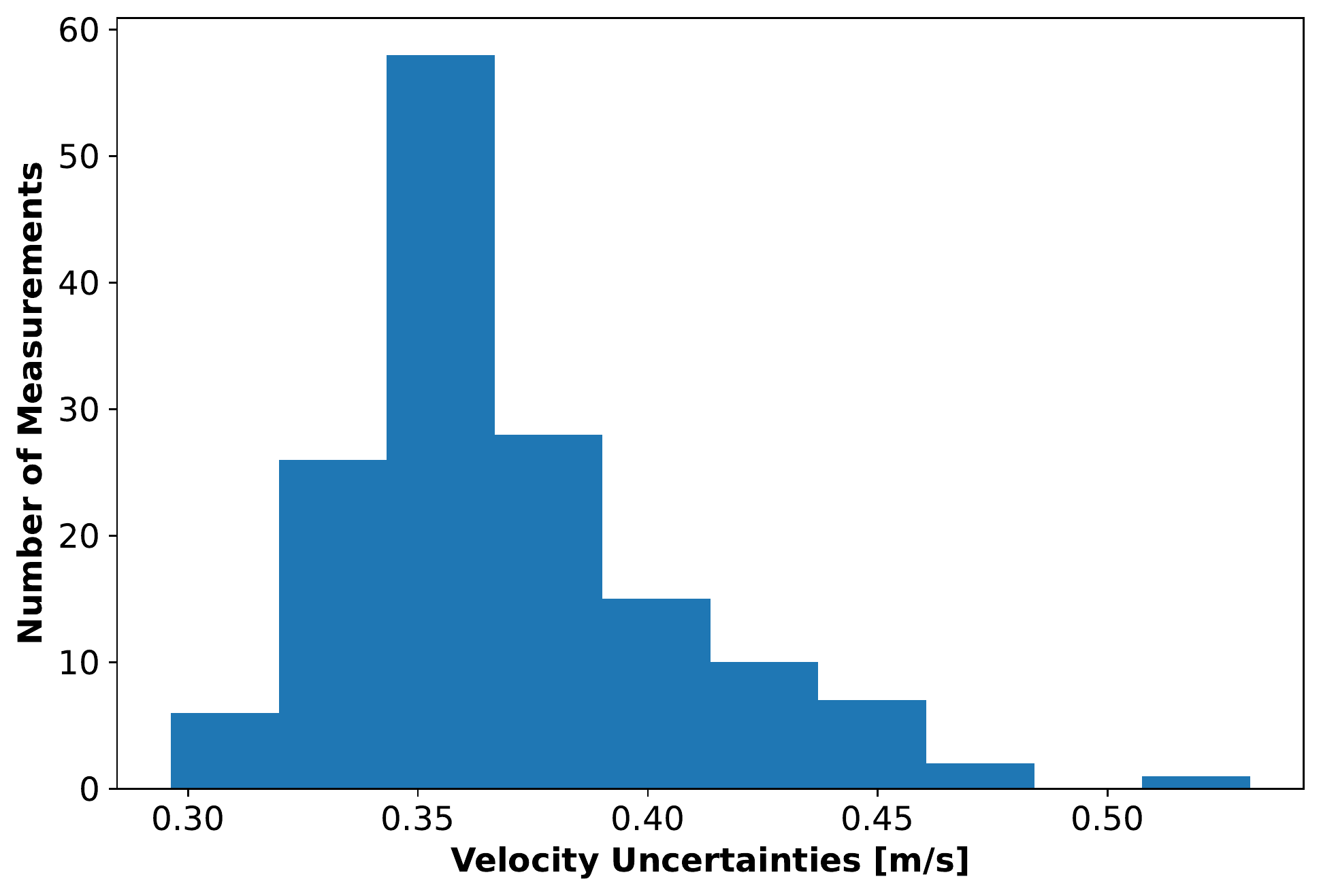}
\caption{Distribution of uncertainties for the \ngoodobs\ observations obtained between May 2020 and May 2023.  The mean uncertainty is 0.37 \ms.  The earliest observations were taken at slightly lower \snr, and cloudy nights or those with bad seeing and correspondingly longer exposure times have slightly increased uncertainties.
\label{fig:uncertainty_dist}}
\end{figure}

\begin{deluxetable}{lrrr}[ht] \label{tab:derived_params}
\tablecaption{ Derived Posteriors }
\tablehead{
  \colhead{Parameter} &
  \colhead{Credible Interval} &
  \colhead{Maximum Likelihood} &
  \colhead{Units}
}
\startdata
  $M_b\sin i$ & $1.093\pm 0.023$ & $1.066$ & M$_{\rm Jup}$ \\
  $a_b$ & $0.2245^{+0.0023}_{-0.0024}$ & $0.2219$ &  AU \\
  $M_c\sin i$ & $28.2\pm 1.5$ & $28.5$ & M$_{\oplus}$ \\
  $a_c$ & $0.4206^{+0.0044}_{-0.0045}$ & $0.416$ &  AU \\
  $M_d\sin i$ & $21.6\pm 2.5$ & $19.5$ & M$_{\oplus}$ \\
  $a_d$ & $0.827\pm 0.011$ & $0.81$ &  AU \\
  $M_e\sin i$ & $3.79^{+0.53}_{-0.54}$ & $3.71$ & M$_{\oplus}$ \\
  $a_e$ & $0.1061\pm 0.0011$ & $0.1049$ &  AU \\\enddata
\end{deluxetable}

\begin{figure*}[ht]
\centering
    \includegraphics[width=0.8\linewidth]{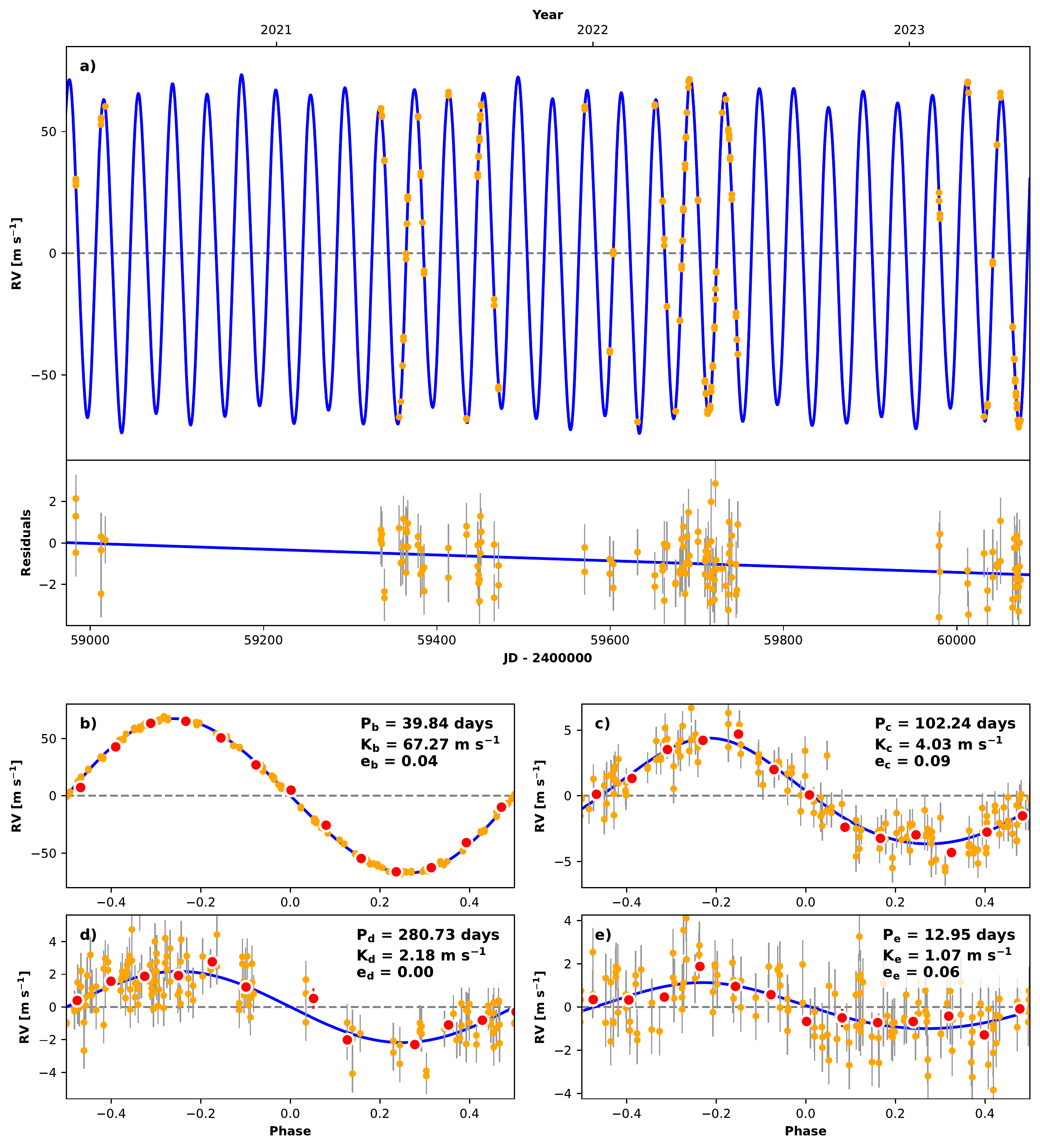}
\caption{Best four planet fit to all EXPRES data with $\snr \ge 200$ with residuals (panel a). The RMS scatter to the residuals of the model to these \ngoodobs\ measurements is 1.20 m/s.  Phased plots for planets b, c, d and e are shown in the panels with those labels.  Orange points in all panels are individual observations with error bars inflated by the stellar jitter term in the model. Larger red points are phase binned.  We still lack complete phase coverage for planet d, since the period is slightly longer than an observing season.  There are only three nights of data from 2020.
\label{fig:4_planet_fit}}
\end{figure*}

The resulting RMS scatter in the residuals to median MCMC model is 1.20 \ms (Figure \ref{fig:4_planet_fit}), which is less than twice that of the quietest star in the \hundredearths\ but 3 times our mean empirically derived single measurement uncertainties for this star of 0.37 \ms (Figure \ref{fig:uncertainty_dist}).  The extra scatter is likely due to stellar activity and possibly additional unresolved planets. \update{We also tested a model with the eccentricity of the fourth planet fixed to zero.  All of the parameters were consistent with the 1-sigma uncertainties; the most notable change was a slightly lower eccentricity for the 102 day planet.  The $\Delta BIC = 2$ shows little support for this model, and the resultant RMS scatter was higher at 1.25 m/s.  We chose to keep the model with the free eccentricity for planet e, but note that additional data will help constrain the eccentricities of all planets.}

\subsection{Dynamical Stability} \label{sec:dynamical}
\begin{figure*}[ht]
\centering
    \includegraphics[width=\linewidth]{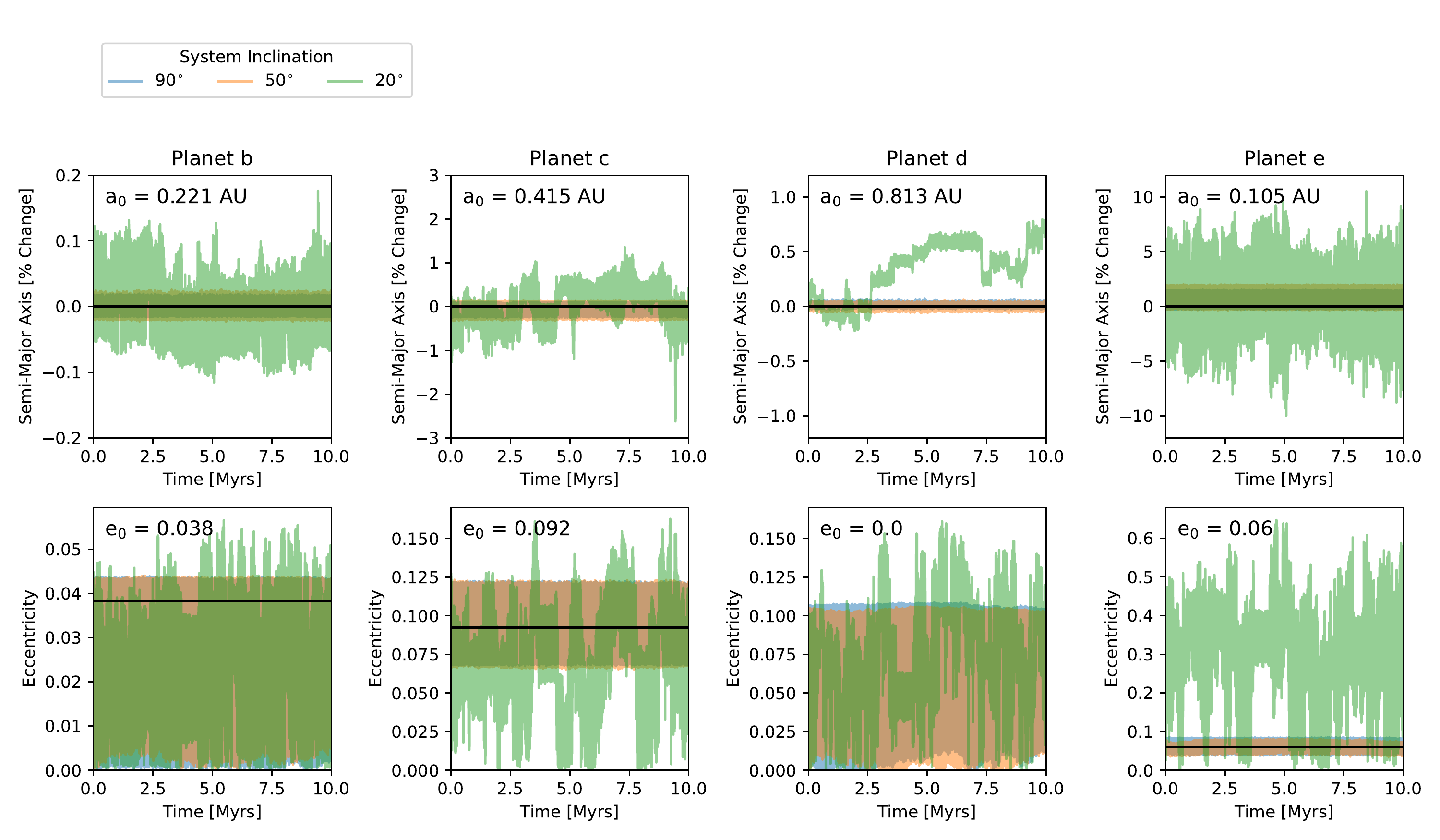}
\caption{Dynamical evolution of the four planets around \rcb.  Each of the four columns corresponds to a planet.  The results of three different dynamical simulations with system inclinations of 90$^{\circ}$ (i.e., edge-on), 50$^{\circ}$, and 20$^{\circ}$ assuming all planets are coplanar are shown in blue, orange, and green, respectively.  The percent change in the semi-major axis of each planet at each time step is shown in the top row; the eccentricity of each planet is shown in the bottom row.
\label{fig:dynamical_simulation}}
\end{figure*}

The system is stable out to at least 10 million years for system inclinations of 90$^{\circ}$ (i.e., edge on orbits) and 50$^{\circ}$ but becomes unstable with a system inclination of 20$^{\circ}$. We ran a dynamical simulation of the system, with the parameters given in Table \ref{tab:orbit_params} and different inclinations, using REBOUND's implementation of the WHFAST integrator \citep{whfast,rebound,reboundwhfast}.  Each time step was of 0.1 days, and we integrated out to 10 million years.  Though there was some change in the orbital parameters for the four different planets, all orbits remained stable with no planets ejected, crossing orbits, or being sent into the host star for 90$^{\circ}$ and 50$^{\circ}$ inclincation.  Figure \ref{fig:dynamical_simulation} shows the evolution of each planet's semi-major axis and eccentricity throughout the 10 million year simulation.

With a system inclination of 90$^{\circ}$ (i.e., edge-on orbits for all planets), the eccentricities of the two lower eccentricity planets, b and d, stay below 0.04 and 0.1 respectively throughout the simulation.  Planet c, with an initial eccentricity of $0.1\pm0.009$, oscillates between 0 and 0.10 in eccentricity.  Planet e's eccentricity ranges between 0.05 and 0.08 over the 10 million year simulation.  The semi-major axes of planets b, c, and d change by much less than 1\% throughout the simulation, while planet e's semi-major axis varies by less than 2\%. With an inclination of 50$^{\circ}$, the situation is nearly identical and the systemremains stable over the 10 million years simulated.  At an incination of 20$^{\circ}$ the system is unstable; all planets experience increased variability in both eccentricity and position. Planet e undergoes changes in semi-major axis of up to 10\% and eccentricity varying beetween 0 and 0.6.

\section{Activity}

Stellar rotation and activity cycles can often mascarade as planetary signals, although high cadence observations mitigate this issue to some extent. Short period planetary signals will necessarily have constant phase; limited spot lifetimes will induce phase variations on the rotational signal.  Although this adds noise to high cadence observations, the phase variations weaken the periodic signal.  In particular, the $\sim13$ day signal is low amplitude and potentially a harmonic of a longer rotation period. We examined activity indicators in the spectroscopic data and analyzed two independent photometric data sets to evaluate the robustness of the identified RV signals.

\subsection{Spectroscopic Activity Indicators}
\begin{figure}[ht]
\centering
\includegraphics[width=\linewidth]{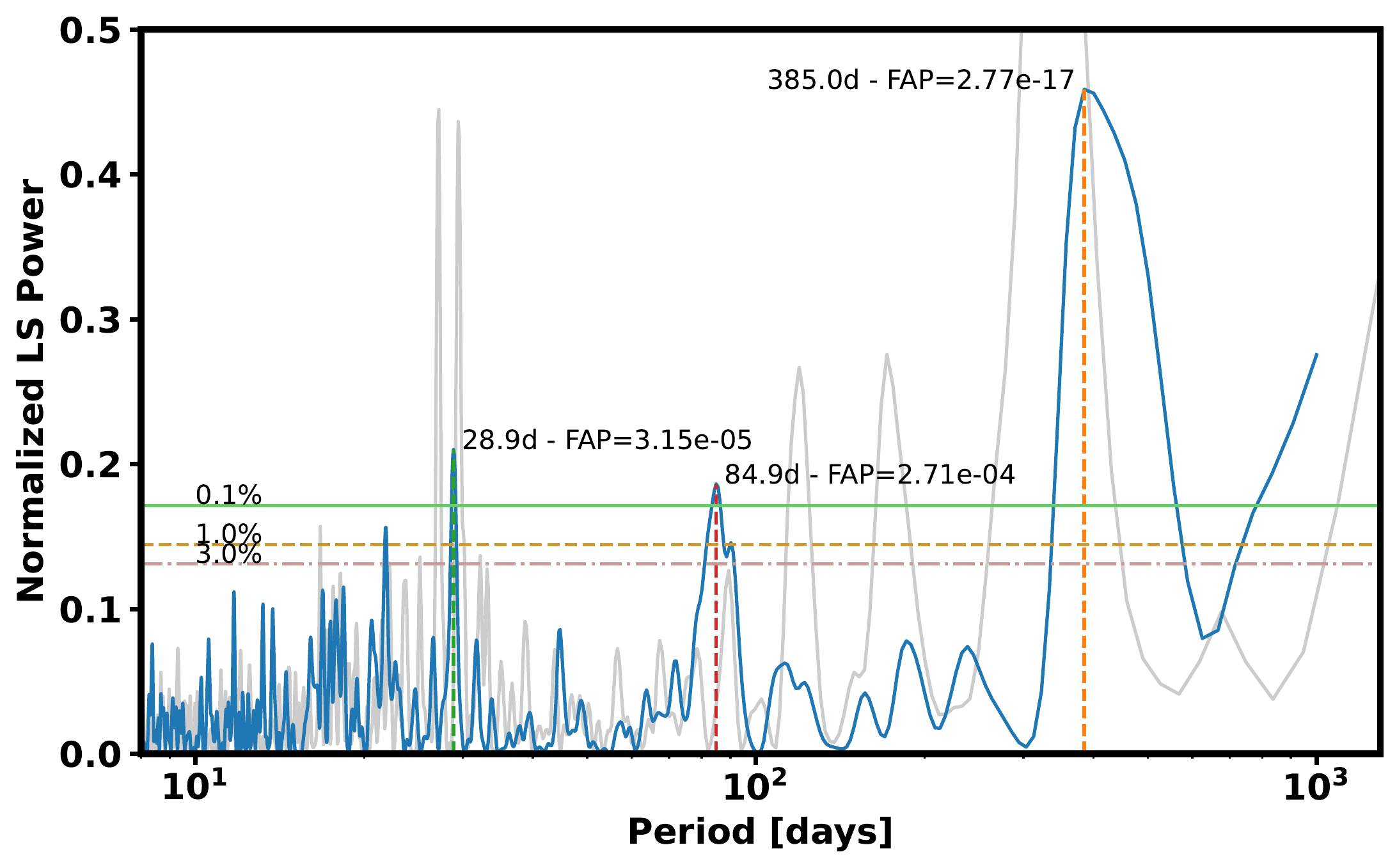}
\caption{The FWHM of the CCF is a measure of line shape changes typically associated with stellar activity.  A Lombe-Scargle periodogram of the CCF FWHM (blue) shows several significant periods \update{that do not coincide with the window function (grey)}, including one at 28.9 days with a possible order 3 harmonic at 84.9 days \citep{10.3847/1538-4365/aab766}.  None of the other activity indicators we measured had any significant periodicity. The 28.9 day period could indicate the rotation period of the star; it is also close to window function periods produced by avoiding observations close to the Moon. Vertical lines indicate periods with false-alarm-probabilities (FAPs) $< 0.1\%$ and are labelled in the legend.  Horizontal lines indicate FAPs of 3\%, 1\%, and 0.1\%.
\label{fig:ccf_fwhm_per}}
\end{figure}

As part of the spectroscopic analysis, we measured the H-$\alpha$ emission, H-$\alpha$ equivalent widths and derived S-values, and during the radial velocity analysis measured the FWHM and bisector span of the cross correlation function.  All of these activivity indicators are obtained for every observation.  Periodogram analysis shows no significant periodicity in any of the indicators with the exception of the FWHM of the CCF (Figure \ref{fig:ccf_fwhm_per}).

The CCF FWHM is a measure of line shape changes typically associated with stellar activity.  We identified \update{three} significant periods at \update{28.9, 84.9, and 385 days.} The 85-day signal is a possible alias of the 29-day signal.  None of the periods we identified coincides with any of the periods in the radial velocities and the period of planet e is not a low-order harmonic of 29 days.  Either of the slowest periods at 28.9 and 84.9 days are reasonable rotation periods for a sub-giant of this mass \citep{10.1051/0004-6361/201219791,10.1088/0004-637x/776/2/67}.  Since the 28.9-day period is a narrower and stronger peak (with an FAP $\sim 10$ times lower), we assume that the longer period is an alias.

\subsection{Photometric Variability}

\update{We searched for rotation signals in both ground based photometry using the T4 0.75m Automated Photoelectric Telescope \citep[APT;][]{henry1999}, and space based photometry from the Transiting Exoplanet Survey Satellite \citep[TESS;][]{ricker2014}.  A detailed discription of the data and analysis can be found in Appendix \ref{appendix:Photometry}.  Due to the 27.4 day observing periods of TESS, we were limited to looking for signals shorter than $\sim 14$ days.  The APT photometry was obtained over 21 observing seasons since 2000 with an average of 54 data points per season.}

\subsection{Stellar Rotation Period Search}
\label{sec:prot}
Rotational periods in the literature for \rcb ranged from 17--20.3 days \citep{baliunas1996,henry2000,10.3847/0004-637x/830/1/46,10.3847/1538-4357/ac1f19}, primarily based on variation in activity indicators.  However, these periods are quite fast for a sub-giant, with typical periods between 30-100 days for stars with mass $< 1.1 M_\odot$ \citep{10.1088/0004-637x/776/2/67}. We performed a period search on the APT data to look for a rotational signature. We found no significant signals (FAP $< 5 \%$) with periods less than 200 days.  The TESS data showed a strong signal at 14.8 days, but this corresponds to a strong peak in the window function; a peak here is expected given the data sampling.

\citet{10.3847/0004-637x/830/1/46} did not find evidence of rotational modulation in photometric data either. They analyzed a light curve from the same APT.  While there is temporal overlap in their data set and the one used here, a new analysis was performed on the observations to provide the light curve included here, as the comparison stars (HD 140716 and HD 144359) used in \citet{10.3847/0004-637x/830/1/46} were found to have low-amplitude, long-term variability.

An old sub-giant is not expected to have many spots that can be used to measure rotation.  Our photometric analysis did not find any clear signals. We do find a strong signal (FAP = $3 \times 10^{-5}$) in the CCF FWHM at 28.9 days (Figure \ref{fig:ccf_fwhm_per}, which could be a rotationally modulated activity signature.  This period would also be consistent with our measured \vsini\ for a differentially rotating star only moderately tilted from a $90^{\circ}$ inclination. Using variations in historical calcium H\&K activity indicators, \citet{10.3847/1538-4357/ac1f19} found a possible rotation period at $20.3 \pm 1.8$ days, with an FAP of 4.3\%. However, their standard spin-down model of rotational evolution predicts a rotation period for \rcb of $52 \pm 5$ days.  They also investigated an alternative model with weakened magnetic braking while on the main sequence.  This results in a period of $28 \pm 2$ days, consistent with the signal we find in the CCF FWHM data.

\section{Discussion} \label{sec:discussion}

The high-cadence and increased precision of the \expres\ \hundredearths\ allows us to identify low-mass and long-period planetary signals.  We have demonstrated that, with less than three years of data, we were able to find the two known planets around \rcb\ as well as two previously undetected planets.  The lower-mass interior planet (e) fits well with those discovered in transit surveys but rarely found in prior RV surveys.  The outer Neptune-mass planet (d) lies in a region of parameter space sparsely populated by either transit or RV surveys due to the prior technical limitations and biases of both.  This opens a window into system architectures that have been missed in prior studies.  In addition, this evolved star is showing conflicting indicators on rotation that could inform our understanding of late stage evolution of planetary systems.

\subsection{Not Peas-in-a-Pod}

Most of the planets in the Solar System orbit within a couple of degrees of orbital inclination of each other, with Mercury being an outlier at ~6$^{\circ}$ away from the orbital plane.  Despite this, the large orbital distances between the planets mean that a transit survey that saw one of our planets would likely miss several others, and the survey would have to run for decades to catch the outer planets.  However, a commonly-found system architecture in the \kepler survey \citep{2011ApJ...736...19B} is a compact system of multiple planets \citep{2015ARA&A..53..409W}.  Subsequent studies have found that, within these systems, the planets tend to have very similar radii \citep{2018AJ....155...48W} and masses \citep{10.3847/2041-8213/aa9714}; the so-called ``peas-in-a-pod''.

Recently, \citet{10.3847/1538-3881/ac7c67} have found that at some point, either the alignment or the tight period spacing breaks down.  The next planet in the pattern is missing, even though that planet should have been detectable given its predicted period and radius.  Their analysis was based on all systems with four or more planets with measured radii. Only four of their 59 systems had detected planets beyond 100 days.

\begin{figure}[ht]
\centering
    \includegraphics[width=\linewidth]{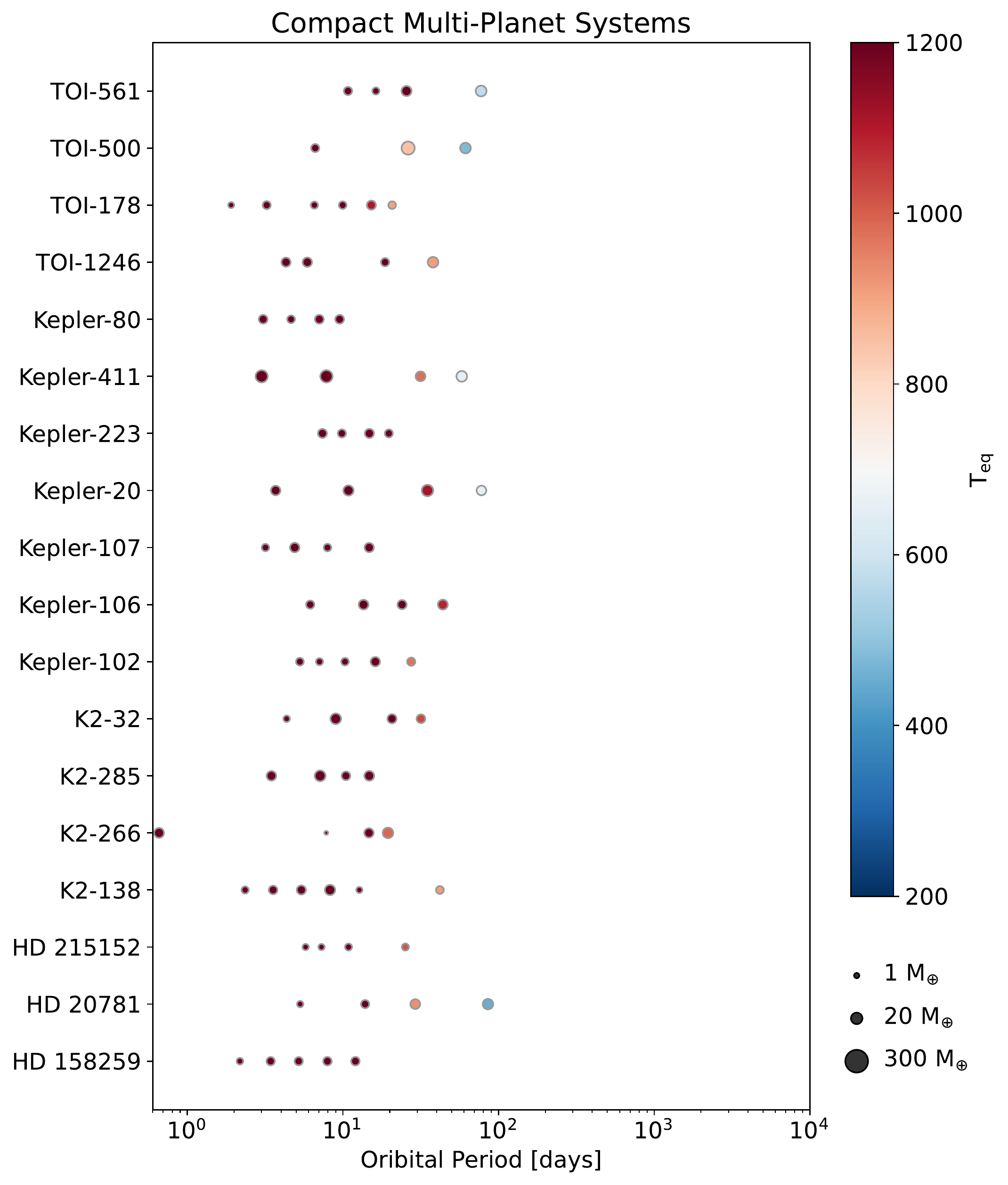}
\caption{Systems with four or more planets with measured masses on orbital periods less than 100 days around G0-M2 stars, from the NASA Exoplanet archive.  The planets within these systems all show a striking uniformity in their mass, as has been noted previously \citep{10.3847/2041-8213/aa9714} for \kepler transiting multi-planet systems.  All but three of these systems were initially detected via transits.  Symbol areas are scaled to planet mass and representative masses are shown in the legend. Symbol colors are scaled to the equilibrium temperatures of the planets.
\label{fig:compact_multis}}
\end{figure}

To place \rcb\ in the context of those systems, we retrieved data from the NASA Exoplanet Archive for comparison.  We have measured minimum planet masses, not radii, so we selected only confirmed systems with four or more planets that have measured masses. In Figure \ref{fig:compact_multis} we plot all those with no detected planets with orbital periods beyond 100 days.  All but three of these systems were discovered via transits, and we see the same types of systems as in \citet{10.3847/1538-3881/ac7c67}, which focused on \kepler compact multis. Additional planets in systems like Kepler-80, Kepler-107, Kepler-223 should have been detectable if the regular period spacing continued.

\begin{figure}[ht]
\centering
    \includegraphics[width=\linewidth]{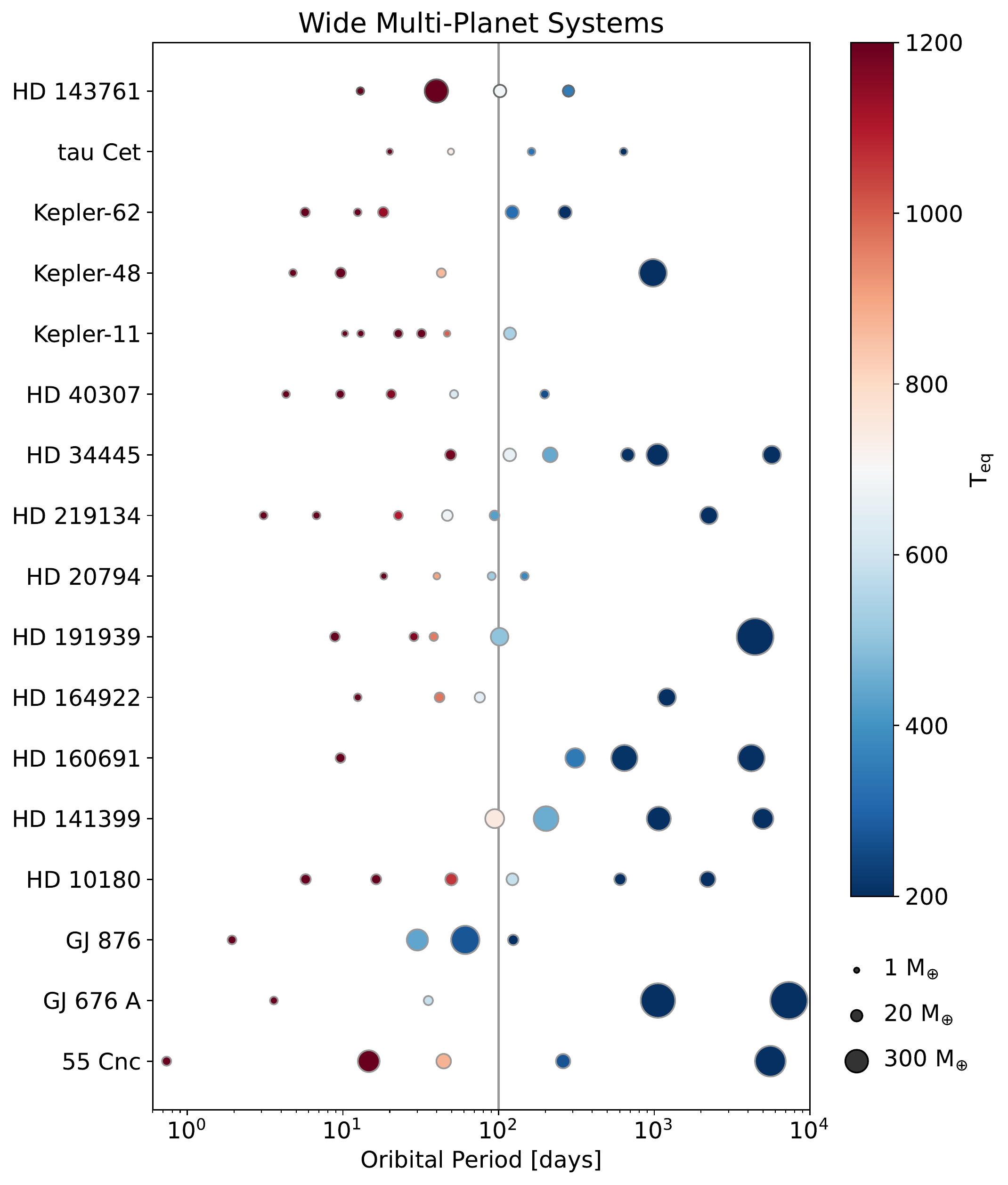}
\caption{Systems with four or more planets with measured masses with at least one having an orbital period $> 100$ days around G0-M2 stars, from the NASA Exoplanet archive.  The systems are less uniform and more widely spaced than those of Figure \ref{fig:compact_multis}, although there are still quite a few ``peas-in-a-pod'' style systems. The topmost system is this one, \rcb. It has small planets both interior to and exterior to a warm Jupiter; a rare trait shared only with GJ 876 and 55 Cnc. Like Figure \ref{fig:compact_multis}, symbol areas are scaled to planet mass, and colors to equilibrium temperatures. Representative masses are shown in the key.
\label{fig:wide_multis}}
\end{figure}

We also selected all systems with four or more planets with measured masses from the Exoplanet Archive that have at least one planet with a period longer than 100 days; \update{these would have been more challenging to detect via transits.} The distributions of planets in these systems look very different (Figure \ref{fig:wide_multis}) than the ``peas-in-a-pod'' systems. There are still some very uniform systems \update{that, inside 100 days, look like the compact multi-planet systems. However, many have a wider dispersion in the masses, and the exterior planets are generally more massive.} The outer planets also break the period spacing pattern of the transit system.

\rcb\ has a warm Jupiter with an interior super Earth \textit{and} two exterior Neptune-mass planets.  A similar scrambling can be seen in GJ 876 and 55 Cnc (Figure \ref{fig:wide_multis}).  Since even 20 \mearth planets such as \rcb d have eluded detection in previous RV surveys, it is possible that there are many more such systems.  These previously overlooked architectures appear more like random assortments of planets than peas-in-a-pod.  EPRV surveys such as the \hundredearths\ run with EXPRES will be able to resolve outstanding questions about the compact multi-planet systems, as well as identify these new more widely spaced architectures.

\subsection{Rotation and v sin i}
\label{sec:vsini}

The full-width-half-maximum (FWHM) of the cross correlation function (CCF) is a measure of spectral line shape variation and is a proxy for stellar activity. We find a probable rotation period for \rcb derived from the CCF FWHM of 28.6 days. This is longer than most previously reported rotation periods for the star, although its low activity signal makes it challenging to obtain a significant signal.  A 28.6 day rotation period is consistent with expectations for a $\sim 10$ Gyr solar mass star and the weakened magnetic braking model of \citet{10.3847/1538-4357/ac1f19} for \rcb specifically.  Our measured \vsini\ of 0.8 km/s then implies that, although this is not a strictly edge-on-orbit, it is not an extremely low inclination.  The equatorial velocity ($v_{eq} = 2 \pi R/P$) would be $\sim 2.37$ km/s; the effects of differential rotation and a modest tilt would result in a low measured \vsini.

\subsection{Formerly Habitable Zone Neptune}

\begin{figure}[ht]
\centering
    \includegraphics[width=\linewidth]{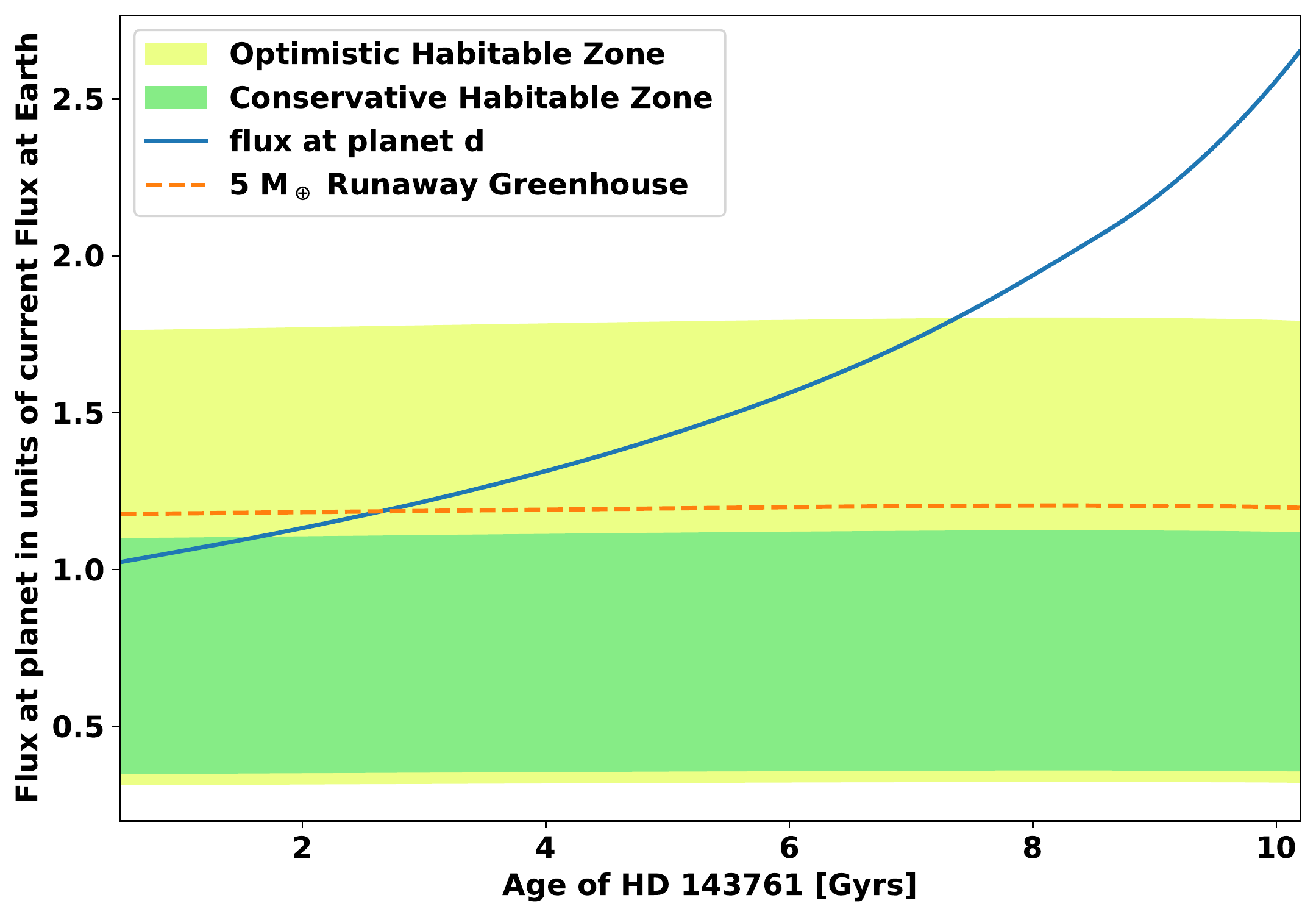}
\caption{Stellar flux at the orbit of planet d, $\sim 281$ days, over the host's main sequence lifetime (blue line).  Also shown are the optimistic (yellow) and conservative (green) habitable zones for a 1 \mearth\ planet over the same time period.  The inner edge of the habitable zone, where a runaway greenhouse would begin, is closer to the host for a 5 \mearth\ planet (orange dashed line).
\label{fig:habitable_zone}}
\end{figure}

Planet d resides in a 281-day orbit around this 0.95 M$_\odot$ star. Although this nine-month period raises hopes of a habitable zone planet, \rcb\ is currently a subgiant with a luminosity more than 75\% greater than the Sun (Table \ref{tab:stellar_props}).  We can calculate the equilibrium temperature for planet d:

\begingroup
  \tiny   
  \thinmuskip=\muexpr\thinmuskip*5/8\relax
  \medmuskip=\muexpr\medmuskip*5/8\relax  
\begin{equation} \label{eqn:teq}
    T_{eq} = \left( \frac{1-A_B}{1-A_{B,\oplus}} \right)^\frac{1}{4}
            \left( \frac{R_*}{R_\odot} \right)^\frac{1}{2}
            \left( \frac{M_*}{M_\odot} \right)^\frac{1}{3}
            \left( \frac{P}{P_\oplus} \right)^{-\frac{2}{3}}
            \left( \frac{T_{\mathrm{eff}}}{T_{\mathrm{eff,\odot}}} \right)
            T_{eq,\oplus}
\end{equation}
\endgroup

The giant planets in the Solar System and Earth all have similar albedos ($\sim 0.3$). Absent additional information we will use Earth's albedo; for G type stars, this is a reasonable assumption \citep{10.3847/1538-4357/ab3be8}. We find $T_{eq} = 345$ K, almost 90 K hotter than the Earth.  However, while the star was on the main sequence, this region would have been much more temperate.  Using the $0.95 M_\odot$ MIST evolutionary track found using our best-fit isochrone, we calculated the stellar flux at the location of the planet over its main-sequence lifetime (Figure \ref{fig:habitable_zone}).  Early in its life, planet d would have resided in the conservative habitable zone, and even out to 7 Gyrs would have been in the optimistic habitable zone \citep{2016ApJ...830....1K}. More massive planets can accomodate a higher flux without entering a runaway greenhouse  \citep{2014ApJ...787L..29K}, as illustrated by the 5 \mearth\ inner edge in Figure \ref{fig:habitable_zone}. Although we might not expect liquid water on the surface of giant planets, Neptune-mass planets in this region are currently rare.

\subsection{An Ancient Metal-poor Exoplanetary System}

\rcb\ is only a mildly metal poor star at its current apparent metallicity of [Fe/H] $= -0.20$.  Over its 10 Gyr lifespan, atomic diffusion has reduced the surface abundances from an initial [Fe/H] $= -0.1$.  In either case, it sits on the low end of the giant planet metallicity correlation \citep{2005ApJ...622.1102F}.  As host to a hot Jupiter, this has made \rcb\ a bit of an outlier from the start.  With at least four companions, this star remains an outlier with its random assortment of planetary masses.  The low metallicity has also meant a slightly hotter environment for its planets, given the stellar mass.

\rcb sits at the edge of evolving into a sub-giant, already having expanded more than 35\% from its original radius.  This is a particularly interesting region to look at the fate of planets near the end of the main sequence.  Planet e is stable on 10 Myr timescales, assuming moderately inclined orbits.  However, each of the planets seems to be moderately eccentric.  Did the evolution of the star onto the sub-giant branch alter the Q of the planets and destabilized the orbits recently?

\section{Conclusions} \label{sec:conclusions}

We have yet to detect planetary systems that look like the Solar System.  These systems have been hidden by lack of RV precision, low cadence, and bias in transit detections.  A primary goal of the \hundredearths\ is to determine the frequency of systems like ours.  Only by fully exploring the parameter space will we be able to reveal the diversity and frequency of planetary system architectures.  \update{There are already several EPRV spectrographs with high cadence exoplanet programs taking science data and several more in commissioning or planned \citep[e.g.][]{10.1117/12.2234411,10.1117/12.2561783,2016SPIE.9908E..6TJ,10.1117/12.2232111,10.1117/12.2629428,10.1051/0004-6361/202038306}.} This work shows that a high cadence survey with an extreme precision spectrograph like \expres\ can begin to give us those answers.  In only a few observing seasons, we recovered two known planets around \rcb\ and identified two new planets: a hot super-Earth and a temperature Neptune mass planet.  The system has an uncommon architecture, hosting a series of close-in planets that have a wide range of masses.

We have had great success over the last 27 years in finding thousands of planetary systems.  Most of those discoveries look nothing like what we expected.  Instead of Solar System analogs, we have found hot Jupiters, sub Neptunes, super Earths, and systems with tightly packed inner planets; none of which exist in the Solar System.  It might seem that systems like ours are rare, but despite the long search, limits in our precision and biases in our detection methods have left large regions of the mass and period parameter space unexamined.  We are now beginning to probe that space.  In these early days, we are not yet finding systems similar our own, but systems like \rcb\  are not like those that are so prevelant in transit surveys either.  Revealing what lies hidden at longer periods can help us better understand planet formation and migration.

\section{Acknowledgements}
These results made use of data provided by the EXPRES team using the EXtreme PREcision Spectrograph at the Lowell Discovery telescope, Lowell Observatory. Lowell is a private, non-profit institution dedicated to astrophysical research and public appreciation of astronomy and operates the LDT in partnership with Boston University, the University of Maryland, the University of Toledo, Northern Arizona University and Yale University. EXPRES was designed and built at Yale with financial support from NSF MRI-1429365, NSF ATI-1509436 and Yale University. Research with EXPRES is possible thanks to the generous support from NSF AST-2009528, NSF 1616086, NASA 80NSSC18K0443, the Heising-Simons Foundation, and an anonymous donor in the Yale alumni community. 

We gratefully acknowledge the support for this research from NASA grant 80NSSC21K0009, NASA XRP-80NSSC21K0571 and NSF 2009528.  RMR acknowledges support from the Heising-Simons 51 Pegasi b Postdoctoral Fellowship. We also acknowledge the generous support for telescope time from the Heising-Simons foundation. G.W.H. acknowledges long-term support from NASA, NSF, Tennessee State University, and the State of Tennessee through its Centers of Excellence Program.

This research has made use of the NASA Exoplanet Archive, which is operated by the California Institute of Technology, under contract with the National Aeronautics and Space Administration under the Exoplanet Exploration Program.

This publication makes use of The Data \& Analysis Center for Exoplanets (DACE), which is a facility based at the University of Geneva (CH) dedicated to extrasolar planets data visualisation, exchange and analysis. DACE is a platform of the Swiss National Centre of Competence in Research (NCCR) PlanetS, federating the Swiss expertise in Exoplanet research. The DACE platform is available at https:\/\/dace.unige.ch.

\appendix
\section{Photometry}\label{appendix:Photometry}

\subsection{TESS Photometry}

\label{sec:tess}
\rcb was observed by the Transiting Exoplanet Survey Satellite \citep[TESS;][]{ricker2014} during Sectors 24, 25, and 51 at 2-minute cadence as well as during Sector 51 at a 20-second cadence. For the analysis discussed in Section \ref{sec:prot}, 2-minute observations are sufficient, so the 20-second cadence observations in Sector 51 are not considered. TESS observing periods are only 27.4 days per sector, limiting our activity analysis with this data to periods shorter than $\sim 14$ days.

\begin{figure}[htb]
\centering
    \includegraphics[width=\linewidth]{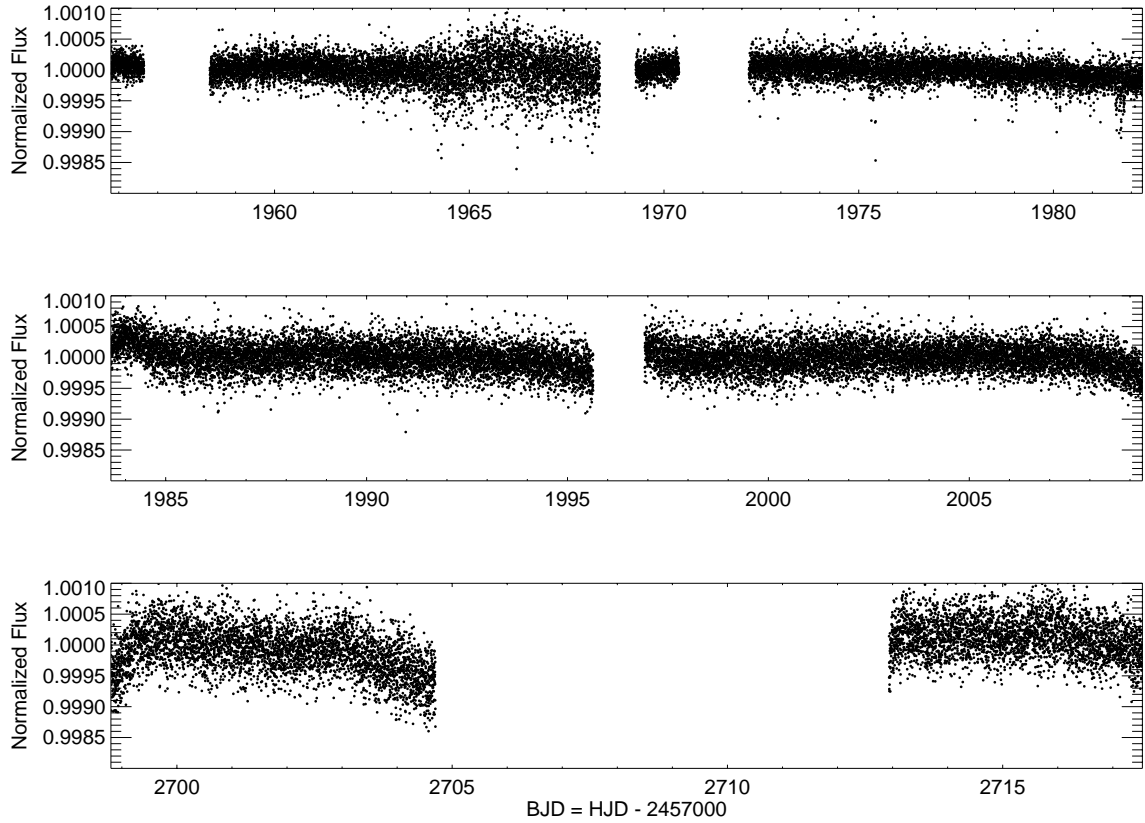}
    \vspace{-1cm}
\caption{TESS light curves from Sectors 24 (top),  25 (middle), and 51 (bottom).  Four CBVs have been removed from each light curve.
\label{fig:tesslcs}}
\end{figure}

We removed cotrending basis vectors (CBVs) from the simple aperture photometry (SAP) light curves.  This process preserves signals from stellar astrophysics while accounting for systematic variations affecting TESS observations.  For the 2-minute cadance observations the first four CBVs were removed.  
Data points with non-zero quality flags were also removed.  From the 2-minute Sector 51 data, data points between 2459712.8167376 and 2459712.9347927, inclusive, were removed due to likely systematic issues that were not addressed by the CBVs or the quality flags. The data with CBVs and flagged data points removed are shown in Figure \ref{fig:tesslcs}.  The data and CBVs were retrieved from the Barbara A. Mikulski Archive for Space Telescopes (MAST).

A periodogram analysis of the TESS light curve was performed combining Sectors 24 and 25.  Sector 51 was not included in this analysis due to the large gap between the sectors and the large data gap in the sector itself. The strongest signal in the periodogram of the TESS data occurs at 14.8 days. However, this corresponds to a strong peak in the window function and is where we would therefore expect to see a signal given the sampling of the data.

\subsection{Automated Photoelectric Telescope Photometry}
\label{sec:APT}

Ground-based photometry of \rcb was obtained at the Fairborn Observatory, AZ with the T4 0.75m automated photoelectric telescope \citep[APT;][]{henry1999}.  Observations were obtained of both \rcb and comparison star HD 139389 in Str\"omgren $b$ and $y$ passbands.  The data were combined into a single passband of $(b+y)/2$, and differential magnitudes for \rcb are included in Table \ref{tab:APT_phot}.  Observations were obtained between 2000 January 13 and 2020 June 26, with an average of 54 data points in each of 21 observing seasons.  

\begin{figure}
\centering
    \includegraphics[width=\linewidth]{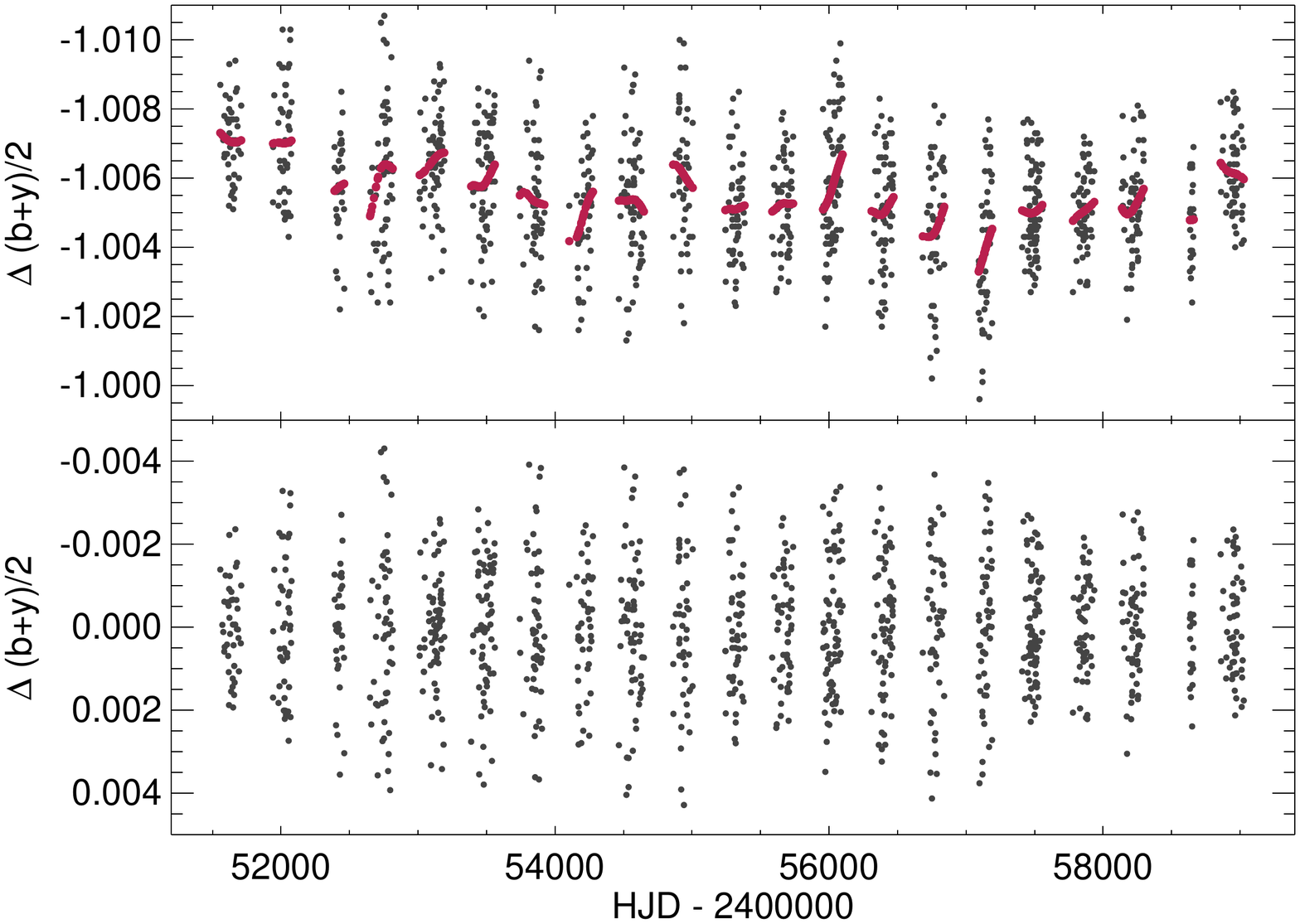}
    \vspace{-1cm}
\caption{Differential photometry of \rcb from the ground-based APT spanning 21 observing seasons.  Top:  $\Delta (b+y)/2$  data (dark gray) with the long-term trend overplotted (red).  Bottom:  The same light curve, with the long-term trend removed. 
\label{fig:apt_phot}}
\end{figure}

A trend was removed from the photometry to account for any signatures that might be long-term, leaving those attributable to the rotation of surface features.  The data were smoothed with a Gaussian kernel with a FWHM of 100 days \citep[following][]{cabot2021,roettenbacher2022}.  The data are included in Figure \ref{fig:apt_phot} and Table \ref{tab:APT_phot}. 

\begin{deluxetable}{ccc}
\tablecaption{APT Photometry
\label{tab:APT_phot}}
\tablehead{\colhead{Heliocentric Julian Date} & \colhead{$(b+y)/2$ Differential} & \colhead{Trend } \\ \colhead{(HJD - 24000000)} & \colhead{Magnitude} & \colhead{ Removed}}
\startdata
51557.0280 & -1.0087 & -1.0073 \\
51572.0254 & -1.0073 & -1.0072 \\
51579.0347 & -1.0071 & -1.0072 \\
51585.9417 & -1.0066 & -1.0072 \\
51586.9829 & -1.0078 & -1.0072 \\
$\cdots$ & $\cdots$ & $\cdots$
\enddata
  \tablecomments{This table is available in machine-readable form.}
\end{deluxetable}

\section{Posterior Parameter Distributions}\label{appendix:Distributions}

We used an MCMC analysis to determine the uncertainties in the parameters.  We used 50 walkers and the first 300,000 steps of burn-in were discarded; the chains were well mixed after 1,920,000 steps.  The eccentricity of planet d was fixed to zero, and the remaining eccentricities were constrained to be less than 0.2.  Gaussian priors were placed on the periods and RV semi-amplitudes of all planets with $\mu$ and $\sigma$ set to the parameters and uncertainties returned from the initial maximum likelihood fit.  The jitter used a Guassian prior with $\mu = 0.35$ and $\sigma = 0.2$.

\begin{figure*}
\centering
    \includegraphics[width=\linewidth]{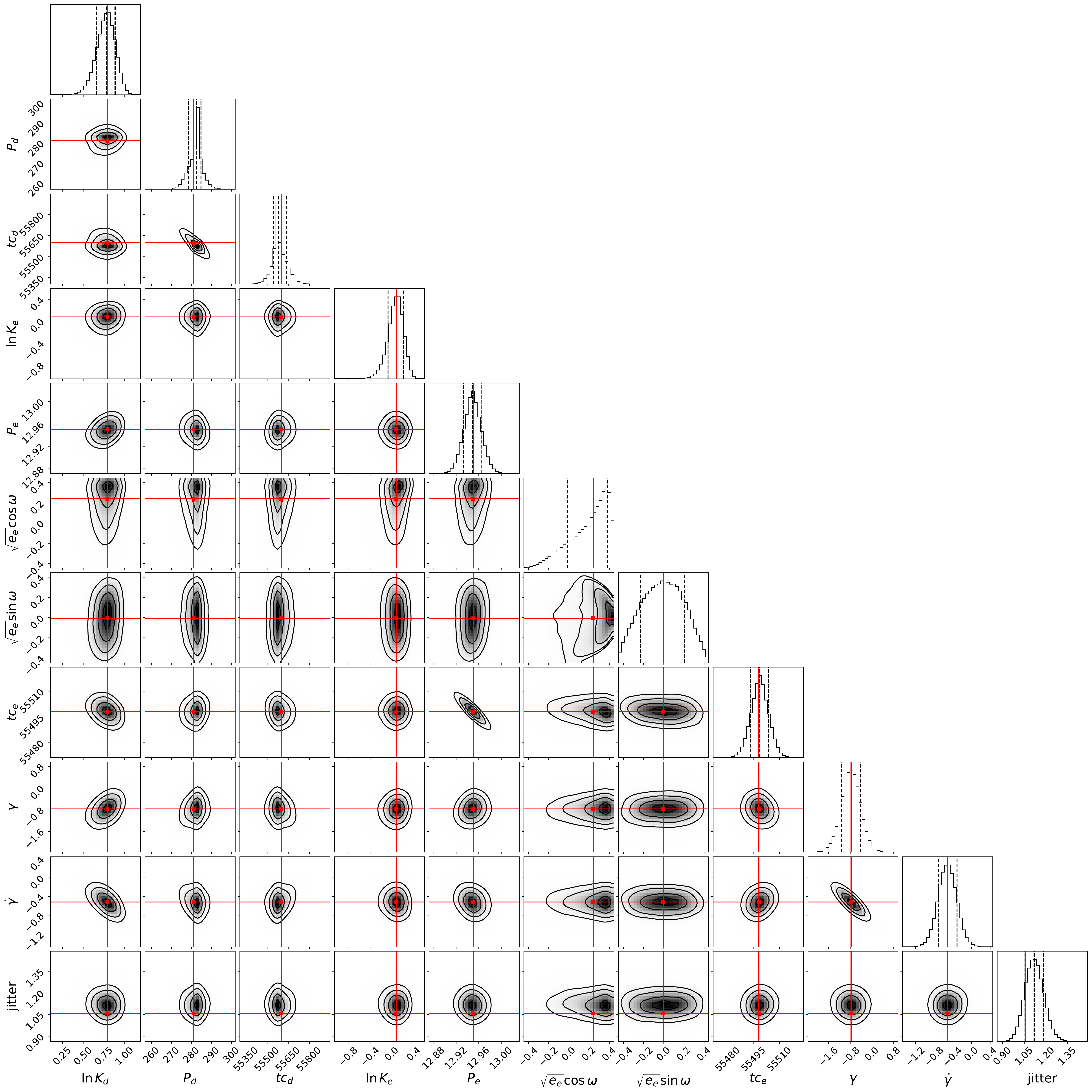}
\caption{Posterior distributions for the model parameters.  The parameter distributions for planets b and c are relatively Gaussian in shape and have been excluded from the plot to increase clarity.  The vertical dashed lines in the single parameter distributions mark the 16th, 50th, and 84th percentiles of the distribution.  The solid red lines are the location of the maximum likelihood values for each parameter in both the 1-D and 2D distributions.
\label{fig:corner_params}}
\end{figure*}

In Figure \ref{fig:corner_params} we plot the posterior distributions for the parameters of planets d and e along with the offset ($\gamma$), linear drift ($\dot{gamma}$), and jitter.  The linear drift has been converted to $m s^{-1} yr^{-1}$ to aid legibility.  The distributions of the parameters for planets b and c are small and nearly Gaussian shaped and were also excluded to aid legibility.

\begin{figure*}
\centering
    \includegraphics[width=\linewidth]{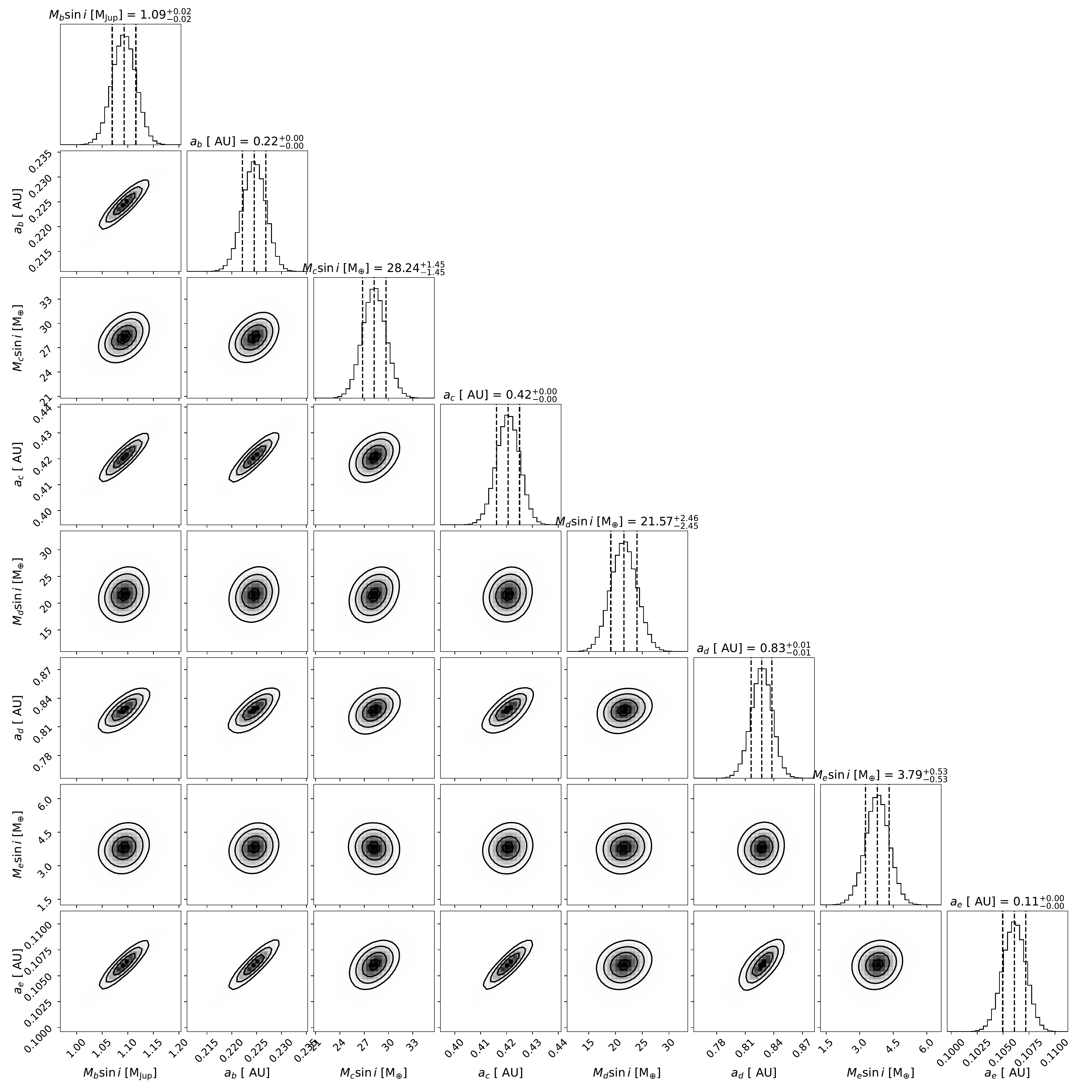}
\caption{Posterior distributions for the derived masses and semi-major axes for planets b, c, d, and e.
\label{fig:corner_derived}}
\end{figure*}

In addition, we have included the posteriors for the derived parameters (Figure \ref{fig:corner_derived}).

\bibliography{expres_iv.bib}{}
\bibliographystyle{aasjournal}



\end{document}